\documentclass{article}
\usepackage{amsfonts,amsmath}
\mathchardef\mhyphen="2D 
\usepackage[a4paper]{geometry}
\usepackage{microtype}

\def\zo/{$0\mkern2mu\mhyphen1$}
\def\Ex{E_{\times}}
\def\nn/{$n \times n$}
\def\a{\alpha}
\newtheorem{conj}{Conjecture}
\DeclareMathOperator\perm{perm}
\title{A Mysterious Cluster Expansion Associated to the Expectation Value of the Permanent of 0-1 Matrices (Revised)}
\date{\today}
\author{Paul Federbush\\
Department of Mathematics\\
University of Michigan\\
Ann Arbor, MI, 48109-1043}
\begin{document}

\maketitle
\begin{abstract}
We consider two ensembles of $n\times n$ matrices.
The first is the set of all $n\times n$ matrices with entries zeroes and ones such that all column sums and all row sums equal $r$, uniformly weighted. The second is the set of $n \times n$ matrices with zero and one entries where the probability that any given entry is one is $r/n$, the probabilities of the set of individual entries being i.i.d.'s. 
Calling the two expectation values $E$ and $E_B$ respectively, we develop a formal relation 
\begin{equation}
\label{A1}
\tag{A1}
E(\perm(A)) = E_B(\perm (A)) e^{\sum_2 T_i}.
\end{equation}
We also use a well-known approximating ensemble to $E$, $E_1$. We prove using $E$ or $E_1$ one obtains the same value of $T_i$ for $i\le20$. (THE PUBLISHED VERSION OF THIS PAPER ONLY
OBTAINS RESULTS FOR $i\le7$. We go beyond the results of the published version by taking much more
advantage of recent work of Pernici and of Wanless on i-matchings on regular bipartite
graphs.)
These terms $T_i$, $i\le20$, have amazing properties. We conjecture that these properties hold also for all $i$. 

\end{abstract}

We happily inform the reader that no knowledge of cluster expansions is necessary to read and understand this paper, but for those interested a general exposition is given in \cite{b}.
The development of this paper is a short sequence of computational steps. 
We will clearly state which steps are rigorous and which are formal.
There will be no theorems, but the formalism has intrinsic beauty. 
Our results will be computer computations (in integer arithmetic) of a large number of terms in the developed expansion, that again have a beauty and compelling force for theorems to be proved.

We begin with the definitions of the objects we will deal with. For an $n\times n$ matrix $A$, in addition to the permanent of $A$, $\perm(A)$, we will need information about the permanents of submatrices of $A$. Given a set of $i$ of the rows of $A$, and a set of $i$ of the columns of $A$, an $i\times i$ submatrix is determined. The sum of the permanents of all such $i\times i$ submatrices of $A$ we denote by $\perm_i(A)$.

We will consider a number of ensembles of $n\times n$ matrices, each with an associated expectation:

The first ensemble is the set of all \zo/ \nn/ matrices with all row and column sums equal to~$r$. 
We let $E$ denote the expectation determined by a uniform weighting in this ensemble.
The expectation $ E(\perm_{m}(A))$ has a natural graph theoretic interpretation. Let $\mathcal{S}$ be the 
set of r-regular bipartite graphs with $2n$ vertices. To each graph $g$ in  $\mathcal{S}$ we associate the cardinality
of its symmetry group $cs(g)$ . We let $n(m,g)$ be the number of m-matchings on $g$. Equivalently,  $n(m,g)$  may
be defined as the number of ways of laying down m dimers on $g$. Then $ E(\perm_{m}(A))$ is the average value
of the ratio $n(m,g)/cs(g)$ over the $g$ in  $\mathcal{S}$.

The second is the Bernoulli random matrix ensemble where each entry independently has a probability $p=r/n$ of being one, and is zero otherwise. 
We denote the associated expectation by~$E_B$.

The third ensemble is again a set of nonnegative integer matrices determined by the second measure in Section~4 of \cite{fkm}.
We denote the associated expectation by $E_1$. $E_1$ is developed using random permutation matrices. Here
we employ permutations on $rn$ objects. Each single such permutation matrix determines an $n \times n$ matrix and the measure
$E_1$ is the uniform measure on the $(rn)!$ $n \times n$ matrices determined by the $(rn)!$ such permutations. Each permutation
is naturally represented as an $rn \times rn$ $0-1$ matrix, $A'$, Entries are said to be in the same residue class if their indices differ by a vector of the form $(an,bn)$ for some integers $a$ and $b$. The residue classes are in $1-1$ correspondence with indices in the $n \times n$ matrix, $B'$, formed of the first $n$ rows and $n$ columns of the $rn \times rn$ matrix. We take as each entry in this determined matrix,
$B'$, the sum of all the entries of $A'$ in the same residue class.

We will later work with a matrix $A$ from the first ensemble and $B$ from the second and then use a product expectation
\begin{equation}
\label{1}
\Ex (f(A)g(B))) = E(f(A)) E_B(g(B)).
\end{equation}

Our initial object of study is the expectation of the permanent in our first ensemble 
\begin{equation}
\label{2}
E(\perm(A)).
\end{equation}
We write $A$ as a sum
\begin{equation}
\label{3}
A = B + ( A - B )
\end{equation}
where $B$ lies in our second ensemble, and we get
\begin{equation}
\label{4}
E(\perm(A))=\Ex (\perm(B+(A-B))).
\end{equation}
We write 
\begin{equation}
\label{5}
A-B\equiv C
\end{equation}
and note that
\begin{equation}
\label{6}
\Ex (\perm(B+C))=\sum_\a E_B(\perm(B_{\bar\a})) \Ex (\perm(C_\a))
\end{equation}
where $C_\a$ denotes some submatrix of $C$ (obtained by a selection of a set of the columns and an equal-sized set of the rows) and $B_{\bar\a}$ is the dual submatrix of $B$ (obtained using the complementary sets of rows and columns).
The sum over $\a$ is over all such submatrices.
Equation~\eqref{6} follows from the definition of the permanent, and the fact that $B_{\bar\a}$ and $C_\a$ are statistically
independent.

We now use the very special properties here that the random variables in $B_{\bar \a}$ are statistically independent  from those
in $C_\a$ , and that the expectations on the right side of \eqref{6} each depend only on the size of the respective submatrices. It follows, from these two very special features, that from \eqref{6} one gets
\begin{equation}
\label{7}
E(\perm(A)) = \sum_{i=0}^n f_i\, E_B(\perm_{n-i}(B)) \Ex  (\perm_i(C)).
\end{equation}
Here we have set $\perm_n(A)=\perm(A)$ (for \nn/ matrices) and $\perm_0(A)=1$. $f_i$ is 1 over the number of distinct $i\times i$ submatrices:
\begin{equation}
\label{8}
f_i = \frac{1}{\binom n i ^2}.
\end{equation}

Equation (7) is derived in Appendix A.
But one may check that \eqref{6} and \eqref{7} are equal, using the special properties above.

We now study $\Ex (\perm_i(C))$. Here we get
\begin{equation}
\label{9}
\Ex (\perm_i(C)) = \sum_{k=0}^i f(i,k)\, E(\perm_{i-k}(A))(-1)^k E_B(\perm_k(B))
\end{equation}
where 
\begin{equation}
\label{10}
f(i,k) = \frac{\binom n i ^2 \binom i k^2}{\binom n k ^2 \binom n {i-k}^2}.
\end{equation}
We note 
\begin{equation}
\label{11}
f_i = f(n,i).
\end{equation}
We derive equation (9) in Appendix A.

We do a little calculation from the easy formula 
\begin{equation}
\label{12}
E_B(\perm_i(B)) = \binom n i^2 i!\, (r/n)^i
\end{equation}
to get 
\begin{equation}
\label{13}
\frac{E_B(\perm_{n-i}(B))}{E_B(\perm (B))} = \frac{1}{r^i} \frac{n^i (n-i)!}{n!} \cdot \frac1{f_i}.
\end{equation}

We put together the above formulas to get our expression for $E(\perm(A))$:
\begin{equation}
\label{14}
E(\perm(A)) = E_B(\perm(B)) \cdot \biggl(1 + \sum_{i=2}^n C_i\biggr)
\end{equation}
with 
\begin{equation}
\label{15}
C_i=
\biggl(
\frac{1}{r^i} \frac{n^i (n-i)!}{n!} 
\biggr)
\cdot \sum _{k=0}^ i f(i,k)\, E(\perm_{i-k}(A)) (-1)^k E_B(\perm_k(B)).
\end{equation}

We have used the fact that $C_1 =0$ (and $C_0=1$). We emphasize so far that everything is ``rigorous'', formulae~\eqref{14} and~\eqref{15} give a neat expression for $E(\perm(A))$.
And also note that all the expectations involving $B$ are easily known by~\eqref{12}.

We now construct our cluster expansion
\begin{equation}
\label{16}
\biggl(
1 + \sum_{i=2}^n C_i
\biggr)
\,\,
\textrm{``}\!=\!\textrm{''}
\,\,
e^{\sum_{i=2}^\infty T_i}.
\end{equation}
This is our first ``formal'' step. 
The $T_i$ are constructed, working with formal power series, by the following
relation
\begin{equation*}
[x^d](1 + \sum_{i=2}^n C_i x^i) = [x^d] (e^{\sum_{i=2}^\infty T_i x^i})
\end{equation*}
As usual $[x^d]$ extracts the coefficient of the $x^d$ term in a (formal) power series.
 We present the first few $T_i$ so that one may see the pattern:
\begin{subequations}
\begin{align}
\label{17a}
T_2 &= C_2
\\
\label{17b}
T_3 &= C_3
\\
\label{17c}
T_4 &= C_4 - \tfrac12 T_2^2
\\
\label{17d}
T_5 &= C_5 - T_3 T_2
\\
\label{17e}
T_6 &= C_6 - T_4 T_2 - \tfrac16 T_2^3 - \tfrac12 T_3^2.
\end{align}
\end{subequations}

As we mentioned, the terms in \eqref{14}, \eqref{15} in the $E_B$ expectations are all known by \eqref{12}.

The theory of Pernici  \cite{p} and Wanless \cite{w} provide a computational
method to compute $E(\perm_m(A))$. This was used in Appendix (C) of
 \cite{p}, but we make much more extensive use herein, leading to our
 much stronger results now than in the published version of this paper. The
 procedure is complicated, and we present it in Appendix B.

$E_1$ is much simpler to compute than $E$, and, as we shall see, is an amazing approximation
for our purposes.
We consider the formulas of \eqref{14}, \eqref{15} as developed using $E_1$ instead of $E$. The formal procedure we followed would have given the same form for \eqref{14}, \eqref{15} with $E$ replaced by $E_1$ . All the expectations using $E_1$ are known from Section 4 of \cite{fkm} as follows
\begin{equation}
E_1(\perm_m(A))  = \binom n m ^2  \frac{r^{2m} m! (rn-m)!}{(rn)!}
\end{equation}

Our computer computations support a few amazing conjectures. In particular
both conjectures hold for $i \le 20$.

\begin{conj}
\label{c1}
Using $E_1$, or $E$ we have that the following limit exists:
\begin{equation}
\label{21}
\lim_{n\to\infty} \frac1n T_i(n).
\end{equation}
\end{conj}

One will realize that the existence of the limit in eq (19) is amazing if one considers that the $C_i(n)$  may behave proportional to $n^{i/2}$ for large $n$ (by computer computation), so much cancellation must take place for the limit in eq (19) to exist.

\begin{conj}
The limits obtained using $E_1$ or $E$ equal each other:
\begin{equation}
\label{21}
\lim_{n\to \infty} \frac 1 n  T_i (n)  \equiv Q_i =\sum_k\frac{a_k(i)}{r^{k}} 
\end{equation}
with $ i-1 \geq k \geq h(i) $, where $h(i) = i/2$ if $i$ is even and $=(i+1)/2$ if $i $ is odd.
\end{conj}

Further if we assume the limits using $E$ and $E_1$ always equal each other
we have calculated, using $E_1$,  the limit $Q_i$ for $i \le 75$, and the form for $Q_i$ as
given in eq (20) holds for these limits.

We present the first few computed $Q_i$, and $Q_{20}$

\begin{equation}
\label{22}
Q_2=\frac{1/2}{ r}
\end{equation}

\begin{equation}
\label{23}
Q_3=\frac{2/3}{ r^2}
\end{equation}

\begin{equation}
\label{24}
Q_4=-\frac{1/2}{ r^2}+\frac{5/4}{ r^3}
\end{equation}

\begin{equation}
\label{25}
Q_5=-\frac{2}{ r^3}+\frac{14/5}{ r^4}
\end{equation}

\begin{equation}
\label{26}
Q_6=\frac{5/6}{ r^3}-\frac{7}{ r^4}+\frac{7}{ r^5}
\end{equation}

\begin{equation}
\label{27}
Q_7=\frac{6}{ r^4}-\frac{24}{ r^5}+\frac{132/7}{ r^6}
\end{equation}

\begin{equation}
\label{28 }
\begin{split}
Q_{20 }=&\frac{1/20}{r^{19}}*(-16796 r^9+2645370 r^8 -68643960 r^7  \\ 
&+686439600 r^6 - 3442004280 r^5  + 9704539845 r^4  - 16087739850 r^3\\
 &      + 15557374800 r^2  - 8119857900 r +  1767263190)
\end{split}
\end{equation}

We would like to consider the convergence properties of the sum of the
$T_i$ in eq (16), but we settle to see what we can find out about the sum
of the limits in $n$, i.e. the sum of the $Q_i$. From the form of the $Q_i$
we might expect this sum converges for r sufficiently large. We look at
the terms in a ratio test for the sum of the $Q_i (r)$. In particular we find
that the ratio of
\begin{equation*}
\frac{Q_i (-1)}{Q_{i -1}(-1)} 
\end{equation*}
to ln(ln(ln(ln(i)))) monotonically increases as $i$ goes from 15 to 75.
This gives a strong suggestion that the sum of the $T_i$ in eq (16)
 diverges.

At the bottom there is a mystery. What theoretical mechanism gives rise to the structure of this cluster expansion?

\textbf{Acknowledgement} We thank Mario Pernici for pointing out some errors in an original form of this paper, as well
as helping with some of the computer calculations.

\section*{Appendix A}
We will derive eq.(9), eq.(7) being a special case of eq.(9). We work in a slightly different setting than in the paper, but 
one having the same computational details. We have a measure space with expectation, $e$. $A$, $B$, and $C$ are
$n \times n$ matrices whose entries are random variables. We assume $A =B + C$ and that the entries of $B$ are statistically
independent of those of $C$. Greek letters label a subset of the indices of the rows and an equal sized subset of the columns.
If there are $r$ elements in each of the subsets determined by $\alpha$, then $A_\alpha$ is an $r \times r $ submatrix of
$A$, and we write $s(\alpha) = r$. We assume
\begin{equation}
e(perm(A_\alpha)) = a ( s(\alpha))
\end{equation}
\begin{equation}
e(perm(B_\alpha)) = b( s(\alpha))
\end{equation}
\begin{equation}
e(perm(C_\alpha)) = c( s(\alpha))
\end{equation}
That is, the expectations of the permanent of a submatrix of a given matrix depends only on the size of the submatrix. We note then
\begin{equation}
e(perm_m(A)) = {\binom n m}^2 a(m)
\end{equation}
\begin{equation}
e(perm_m(B)) =  {\binom n m}^2 b(m)
\end{equation}
\begin{equation}
e(perm_m(C)) = {\binom n m}^2  c(m)
\end{equation}
Analagous to eq.(6) we have that 
 \begin{equation}
perm(B_\gamma +C_\gamma)= \sum_{\alpha \subset \gamma}perm(B_\alpha) perm(C_{\bar\alpha})
\end{equation}
where $\bar\alpha$ is the set of indices of rows and columns inside the sets $\gamma$ complementary to the rows and 
columns of $\alpha$. Taking expectations., and setting $s \equiv s(\gamma)$, we get
\begin{equation}
a(s) = \sum_{i=0}^s \binom s i ^2 b(i) c (s-i).
\end{equation}
Substituting (28)-(33) into (34) we obtain eq(9).
\section*{Appendix B}
We let $A$ be in our first ensemble, i.e. a 0-1 matrix with each row sum and each
column sum equal $r$. We associate a bipartite r-regular graph to $A$. $g$. Its
black vertices are labelled by the column indices and its white vertices by the
row indices. There is an edge connecting a black vertex $j$ with a white
vertex $i$ if and only if the $ij$ entry of the matrix equals $1$. We note that
with this correspondence one has that the number of j-matchings of $g$
is exactly $\perm_j(A)$, that we now call $m_j$. Wanless
developed a formalism to compute the $m_j$ of any regular
graph. We here only give a flavor of this formalism, giving the consequences we will use in this paper.

For each $j \ge 4$ there are defined a set of graphs
$g_{j1},g_{j2},\ldots,g_{j n(j)}$.
Given a regular graph $g$, one computes for each $g_{jk}$ the number of subgraphs of
$g$ isomorphic to $g_{jk}$, call this $g \slash g_{jk}$.
Then $m_i$ for $g$ is determined by the $\sum_{j=4}^{i} n(j)$ values of $g \slash g_{jk}$.
We define $M_i$ to be the value of $m_i$ assigned to any graph with all of these $\sum_{j=4}^{i} n(j)$
values of $g \slash g_{jk}$ zero. Such graphs will exist only for large enough n.
Initially $M_i$ is defined only for such n. But it may be extended as a finite
polynomial in $\frac{1}{n}$ to all non-zero $n$.
$M_i = M_i(r,n)$ is an important object of study to us.
\par In \cite{p} Pernici systematized the results of Wanless. We now 
present the very non-trivial computational construction of $M_i$ from \cite{p} and \cite{w}. One first defines quantities $u_s(r), s \geq 2$ by 
\begin{equation}
	T_r = \frac{2(r-1)}{2(r-1)-r+r\sqrt{1-4x(r-1)}}. \label{eq:3.1} 
\end{equation}
\begin{equation}
	u_s(r) = [x^s]T_r  \label{eq:3.2}.
\end{equation}
The notation is slightly changed from \cite{p}. The expression $[x^s]f$ for a series, $f$, in $x$ is defined as the coefficient of $x^s$ in the series $f$. Then one has
\begin{align}
	M_j = [x^j] \exp{\left(nrx - \sum_{s \geq 2}\frac{nu_s(r)}{s} (-x)^s\right)}\label{eq:3.3}.
\end{align}

Defining quantities $a_h(r,j)$ one has the expressions 
\begin{align}
	M_j &= \frac{n^jr^j}{j!}(1+H_j) \label{eq:3.4}\\
	H_j &= \sum_{h = 1}^{j-1} \frac{a_h(r,j)}{n^h}\label{eq:3.5}
\end{align}
that exhibit the structure of $M_j$ especially in so far as powers of $n$. It is important to keep in mind often that 
\begin{align}
	a_0 &= 1 \label{yes}\\
         a_h &= 0 \text{ if } h \geq j \label{no}
         \end{align}
In \cite{p}, Pernici via a clever formal computation (not rigorous) derives the equations
\begin{align}
   [j^k n^{-h}]\,\ln \biggl( 1 + H_j \biggr) =
    [j^k n^{-h}]\,\ln \biggl( 1 + \sum_{s=1}^{j-1} \frac{a_s(r,j)}{n^s} \biggr) &=
  0, \qquad k \ge h+2 \label{eq:3.6}\\
  [j^{h+1} n^{-h}]\,\ln \biggl( 1 + H_j \biggr)=  [j^{h+1} n^{-h}]\,\ln \biggl( 1 + \sum_{s=1}^{j-1} \frac{a_s(r,j)}{n^s} \biggr)
  &= \frac{1}{(h+1)h} \biggl( \frac{1}{r^h} - 2 \biggr) \label{eq:3.7}
 \end{align} 
equations (16) and (17) of \cite{p}. Here $[j^an^{-b}]f$ is an obvious generalization of $[x^s]f$. The status of these equations (\ref{eq:3.6}),(\ref{eq:3.7}) is as follows. First, (\ref{eq:3.6}) is true and in fact we have a stronger result.
	Equation (\ref{eq:3.6}) holds if $M_j$ is calculated from (\ref{eq:3.3}) with any values for the $u_s$, $s\geq 2$, not necessarily the values given by (\ref{eq:3.1}),(\ref{eq:3.2}), \cite{pf}. 	For $r \leq 10$ and $j \leq 100$ equation (\ref{eq:3.7}) holds, \cite{p}. Further it is also true for
$ j < 30$, all $r$ , see \cite{arc}, both results by computer computation. We also
will use:
\begin{equation}
  [j^k n^{-h}]\,\ln \biggl( 1 + H_j \biggr) =
    [j^k n^{-h}]\,\ln \biggl( 1 + \sum_{s=1}^{j-1} \frac{a_s(r,j)}{n^s} \biggr) =0
  , \qquad k \le h \label{eq:3.8}\\
\end{equation}

We turn to the computation of $M_j$. We use (\ref{eq:3.4}),(\ref{eq:3.5}) expressing
$M_j$ in terms of the $a_h (r,j)$. We compute the $a_h (r,j)$ inductively
in $h$, starting with $h=0$ in (\ref{yes}). We assume $a_0$, $a_1$...
$a_{h-1}$ known, then we determine $a_h$ using (\ref{eq:3.6})-(\ref{eq:3.8}).
This procedure we have implemented by the simplest Maple program. We
plan  in the future to compare this to the computational technique in \cite{p}. 

Returning to the considerations of the first two paragraphs of this appendix,
we note that Wanless in \cite{w} introduces for each $i \ge 4$ an $\epsilon_i$
which is a linear sum of the $n(i)$  $g \slash g_{ik}$. For $i \le 7$ they
are as follows:

\begin{equation}
\begin{aligned}
\epsilon_4 =& 8C_4 \\
\epsilon_5 =& 80\kappa C_4 \\
\epsilon_6 =& 528\kappa^2C_4 + 12C_6 - 48\theta_{2,2,2} \\
\epsilon_7 =& 2912\kappa^3C_4 + 168\kappa C_6 - 672\kappa \theta_{2,2,2}
-56 \theta_{3,3,1}
\end{aligned}
\label{eps4}
\end{equation}
Here $n(4)  = n(5) =1$, $n(6) = 3$, and $n(7) = 4$. $C_4$ is the number of
ways of embedding a 4-cycle into g. (So $g_{41} =g_{51}$, a 4-cycle, and
$C_4 = g \slash g_{41}$. etc. ) $C_6$ is the number of ways of embedding
a 6-cycle into $g$. We now consider a graph with 5 vertices and 6 edges
with two distinguished vertices connected by three distinct paths each of
two edges. $\theta_{2,2,2}$ is the number of ways of embedding this graph
into $g$. Similarly $\theta_{3,3,1}$ is the number of ways of embedding a
graph with two distinguished vertices connected by three paths of lengths
3,3, and 1 edges respectively. We will see that all the detailed information 
in the equations for the $\epsilon_{i}, i\ge 4$ doesn't effect the results we
are after. They are here presented merely to illuminate the rich theory we are
taking advantage of.

Now that the $a_h (r,j)$ are known, we also know $M_j$ by 
eqs \ref{eq:3.4}-\ref{eq:3.5}. We then may find $m_j$ from
\begin{equation}
m_j = \exp(\sum_{s \ge 3} \frac{\epsilon_s}{2s} (-\hat x)^s) M_j
\label{Mim}
\end{equation}
where 
\begin{equation}
\hat x M_j \equiv M_{j-1}.
\end{equation}
See equation (11) in \cite{p} . This expresses $m_j$ as a linear combination
of monomials in $\{ \epsilon(k), k \le j \}$. The coefficients in this expression
are functions of $r$, $j$, and $n$. Following the development from eq (1)
through eqs (17), using our expression for $m_j$, a simple maple program converts $T_j$ /n into a similar linear combination in monomials in $\{ \epsilon(k), k \le j \}$ with coefficients functions of $r$, $j$, and $n$. Results in [5] and [6]
show that the expectations over the ensemble $E$ of the set of our finite set of monomials in the variables
 $\{ \epsilon(k), k \le j \}$ are uniformly bounded in n. And using this it amazingly
 turns out that Conjectures 1 and 2 are true for $i \le 20$ ! \emph{Mirabile dictu}, excuse
 the supine. We can not help but expect this true for all i...a true but unprovable
 result?

\end{document}